\documentclass[12pt,preprint]{aastex}

\def\etal       {et al.}
\def\ie         {{i.e.},\ }
\def\eg         {{e.g.},\ }

\def\AIPS      {{\small AIPS}}
\def\IMAGR     {{\small IMAGR}}
\def\ROBUST    {{\small ROBUST}}

\def\hho       {H$_2$O}

\def\sioone    {SiO v$=1$, J=1--0}

\def\OrionI    {Orion--I}
\def\OrionKL   {Orion--KL}
\def\IRctwo    {IRc~2}
\def\Msun      {M$_\odot$}
\def\Lsun      {L$_\odot$}
\def\Hminus    {H$^-$ free-free}
\def\pe        {${\rm p}^+/{\rm e}^-$}
\def\mjyb   {\ifmmode {{\rm mJy~beam}^{-1}} \else{mJy~beam$^{-1}$}\fi}
\def\jyb    {\ifmmode {{\rm Jy~beam}^{-1}} \else{Jy~beam$^{-1}$}\fi}

\def\Vorb    {\ifmmode {V_{orb}} \else ${V_{orb}}$ \fi}
\def\Vdot    {\ifmmode {\.V} \else ${\.V}$ \fi}

\def\arcm{\ifmmode {' }\else $' $\fi}
\def\arcs{\ifmmode {'' }\else $'' $\fi}
\def\arcmper{\ifmmode \rlap.{'} \else $\rlap{.}' $\fi}
\def\arcsper{\ifmmode \rlap.{''} \else $\rlap{.}'' $\fi}
\def\porm   {\ifmmode\pm\else$\pm$\fi}
\def\kms    {\ifmmode{{\rm ~km~s}^{-1}}\else{~km~s$^{-1}$}\fi}
\def\peryr  {\ifmmode{{\rm ~yr}^{-1}}\else{~yr$^{-1}$}\fi}
\def\percmcube   {\ifmmode{{\rm ~cm}^{-3}}\else{~cm$^{-3}$}\fi}
\def\percmsquare {\ifmmode{{\rm ~cm}^{-2}}\else{~cm$^{-2}$}\fi}
\def\masy   {\ifmmode{{\rm mas~y}^{-1}}\else{mas~y$^{-1}$}\fi}
\def\micron {\ifmmode{\mu{\rm m}}\else{$\mu$m}\fi}
\def\a      {\ifmmode {\rlap.}^{''}\! \else ${\rlap.}^{''}\!$\fi}

\newbox\grsign \setbox\grsign=\hbox{$>$} \newdimen\grdimen \grdimen=\ht\grsign
\newbox\laxbox \newbox\gaxbox
\setbox\gaxbox=\hbox{\raise.5ex\hbox{$>$}\llap
     {\lower.5ex\hbox{$\sim$}}}\ht1=\grdimen\dp1=0pt
\setbox\laxbox=\hbox{\raise.5ex\hbox{$<$}\llap
     {\lower.5ex\hbox{$\sim$}}}\ht2=\grdimen\dp2=0pt
\def\gax{\mathrel{\copy\gaxbox}}

\slugcomment{2007 April 3}

\shorttitle{The High-Mass Protostar \OrionI}

\shortauthors{Reid \etal      }

\begin{document}

\title{ Imaging the Ionized Disk of the High-Mass Protostar \OrionI }

\author{M. J. Reid}
\affil{Harvard--Smithsonian Center for Astrophysics,
    60 Garden Street, Cambridge, MA 02138}
\email{reid@cfa.harvard.edu}

\author{K. M. Menten}
\affil{Max-Planck-Institut f\"ur Radioastronomie,
       Auf dem H\"ugel 69, D-53121 Bonn, Germany}
\email{kmenten@mpifr-bonn.mpg.de}

\author{L. J. Greenhill}
\affil{Harvard--Smithsonian Center for Astrophysics,
    60 Garden Street, Cambridge, MA 02138}
\email{lgreenhill@cfa.harvard.edu}

\author{C. J. Chandler}
\affil{National Radio Astronomy Observatory,
    P.O. Box 0, Socorro, NM 87801}
\email{cchandle@nrao.edu}

\begin{abstract}
We have imaged the enigmatic radio source-I (\OrionI) in 
the Orion-KL nebula with the VLA at 43~GHz with 34~mas
angular resolution.  The continuum emission is highly
elongated and is consistent with that expected from a nearly 
edge-on disk.  The high brightness and lack of strong molecular
lines from \OrionI\ can be used to argue against emission from 
dust.  Collisional ionization and \Hminus\ opacity, as in Mira 
variables, require a central star with $\gax10^5$~\Lsun, which 
is greater than infrared observations allow.  However, if 
significant local heating associated with accretion occurs,
lower total luminosities are possible.  Alternatively, photo-ionization 
from an early B-type star and \pe\ bremsstrahlung can explain our 
observations, and \OrionI\ may be an example of ionized accretion 
disk surrounding a forming massive star.  Such accretion disks
may not be able to form planets efficiently.
  
\end{abstract}

\keywords{infrared: stars --- ISM: individual (Orion Kleinmann-Low) ---
stars: individual (I,IRc 2) --- stars: formation --- 
planetary systems: formation}

\section{Introduction}

     The nearest massive star forming region, the 
Kleinmann-Low (KL) nebula in the Orion molecular cloud,
is at a distance 480~pc \citep{G81}.
While the brightest near-infrared source in \OrionKL\ \citep{KL67} 
is the Becklin-Neugebauer (BN) object, it contributes only
a small fraction of the total nebular luminosity of 
$\sim10^5$~\Lsun\ (see Thronson \etal\ 1986 and references therein).
Other young stellar objects (YSOs) are deeply embedded and hidden
from view at infrared (IR) wavelengths.   
An object giving rise to strong mid-IR emission, \IRctwo, 
has been long suspected to be the dominant energy
source in \OrionKL\ \citep{D81,G81,S04}, but hidden by
$>60$~mag of visual extinction \citep{Gez98,G04}.  
However, \IRctwo\ breaks-up into several compact regions 
\citep{D93,G04}.  Moreover, \citet{Gez92} and \citet{MR95}
showed that \IRctwo\ is offset from the compact radio source-I
(hereafter \OrionI), and some components of \IRctwo\ could arise from 
reflected light.  \OrionI\ is very deeply embedded and
\citet{G04} estimate optical depths $>300$ at 8 and 22 $\mu$m wavelengths.

      The radio source \OrionI\ is an enigmatic object.  
Proper motion measurements suggest that \OrionI\ might have been part 
of a multiple system that disintegrated $\approx 500$ years ago 
\citep{G05}.
Strong \hho\ and OH \citep{C06} masers are concentrated 
near \OrionI, and the \hho\ masers are distributed in an elongated 
pattern along position angle (PA) $\approx45^\circ$ (East of North).  
The \hho\ masers seem to be expanding about a central position in the 
general vicinity of \OrionI\ \citep{G81}.  
\OrionI\ also displays strong SiO masers, which
are usually associated with evolved asymptotic giant branch (AGB) 
or supergiant stars and are very rare in star forming regions \citep{H85}. 
Interferometric maps of the SiO masers with the BIMA
array at an angular resolution of $\approx1\arcs$
indicated the possibility that they came from a rotating and 
expanding disk \citep{P95}.  
Higher resolution observations with the VLA \citep{MR95}
located \OrionI\  at the center of the SiO masers.  Menten \&\ Reid 
argued that the presence of vibrationally-excited SiO masers,
which require temperatures exceeding 1000~K (\eg  Lockett \& Elitzur 1992) 
at a radius of $\approx50$~AU from \OrionI, was strong evidence 
that it must be a very luminous object ($\sim10^5$ \Lsun). 

Observations by \citet{G98} with the VLBA at a resolution of 
$\sim1$~mas clearly resolved the SiO masers into four ``arms''
that together make an 0.2\arcs\ ``X''-like pattern.
Since an X-like pattern has two symmetry axes,
two models for the SiO masers have been forwarded.
The SiO masers could form in the limbs of a high velocity
bi-conical outflow projected along a NW--SE axis \citep{G98,D99}.  
In this model, the
western (eastern) arms are moving away from (toward) the observer and 
hence are red-shifted (blue-shifted) with respect to the systemic velocity.
Alternatively, the SiO masers could form in material being expelled from
a rotating disk, whose spin axis is projected NE--SW.  
In this model, the western (eastern) arms are moving away from (toward) 
the observer owing to rotation \citep{G03}.  As discussed in \S 3, 
evidence favors the latter model of a rotating, nearly edge-on, disk 
centered on \OrionI. 

      We have used the VLA at 43 GHz and imaged the radio continuum
emission associated with \OrionI\ and located it 
precisely toward the center of the SiO maser X-like pattern. 
These data together give the clearest picture yet presented of the 
nature of a high-mass, YSO, disk on scales of $\approx16$~AU.  
In this paper we concentrate on understanding the properties and nature 
of the continuum emission from the disk-like structure of \OrionI.

\section{Observations \& Results}

Our observations were made on 2000 November 10
with the NRAO\footnote{The National Radio Astronomy Observatory
(NRAO) is operated by Associated Universities, Inc., under a cooperative
agreement with the National Science Foundation.}
Very Large Array (VLA) in a manner similar to those of \citet{MR95}.
All 27 antennas had 40--50 GHz receivers, compared to our previous 
observations with only 9 antennas.  The recently installed receivers 
had better noise performance than the ones available 
in 1994 and were placed on antennas providing longer interferometer
baselines. Thus, the present data yielded lower noise levels 
and higher angular resolution compared to our previous data.

In order to image the continuum emission from \OrionI\ and to
locate this emission with respect to the SiO masers,
we employed a dual-band continuum setup with a narrow band (1.56 MHz), 
covering red-shifted \sioone\ masers between LSR velocities of 
10.8 -- 21.3 \kms\ (assuming a rest frequency of 43122.08 MHz), 
and a broad band (50 MHz), centered at 43164.9 MHz
on a line-free portion of the spectrum.  Both frequency
bands were observed in dual-circular polarizations.
We observed from $0^{\rm h}30^{\rm m}$ to $10^{\rm h}30^{\rm m}$
local sidereal time. 
Absolute flux density calibration was obtained from an observation of 
3C286, assuming the flux density spectrum of \citet{B77}.
Observations of the quasar 0501--019, measured to be 0.74~Jy, 
were interspersed with \OrionI\ to monitor gain variations and
to determine electronic phase-offsets between the bands.

The narrow-band data were then ``self-calibrated'' with the 
very strong maser signal as a phase reference.  The phase and 
amplitude corrections were then applied to the broad-band data, 
and a high quality map of the continuum emission was produced. 
A detailed description of this cross-calibration procedure can be 
found in \citet{RM97}.  Once the continuum signals were 
``cross-self-calibrated'' with the SiO maser signals, the data were
imaged with the Astronomical Image Processing System (\AIPS) task
\IMAGR.  We produced maps with two different weightings of the 
(u,v)-data, shown in Fig.~\ref{fig:maps}.  
Using \IMAGR\ weighting parameter ``\ROBUST=5'' the
dirty beam was $58\times45$~mas at a PA of $-20^\circ$, 
and we restored the image with a round beam of 50~mas
(approximately the geometric-mean size). Using ``\ROBUST=0'' the
dirty beam was $41\times28$~mas at PA of $-30^\circ$, and
we restored the image with a round beam of 34~mas.
At a distance of 480~pc, 34~mas corresponds to 16~AU.
These maps had rms noise levels of 0.13~\mjyb and 
0.14~\mjyb, respectively.

In order to image the SiO maser emission at high spectral resolution, 
several scans in spectral-line mode were interspersed with the
dual-band continuum observations.   We covered all of the SiO maser
emission with a bandwidth of 6.25 MHz and 128 spectral channels, 
which provided a velocity resolution of 0.34 \kms.  The line data
were self-calibrated by choosing a channel with strong emission
as a reference, and the resulting phase and amplitude
corrections were applied to the other channels.  Scans of the
strong extragalactic continuum sources 3C~84 and 3C~273
provided bandpass calibration. 
We produced a spectral-line data cube, which we
restored with a 30 mas beam, taking advantage of the high
signal-to-noise ratio and slightly ``over-resolving'' the
dirty beam of $43\times27$ mas.

Alignment of the continuum and maser emission to about 
5~mas accuracy was achieved by producing a 
pseudo-continuum map from the line data, using spectral channels 
covering the velocities that were within the 1.56 MHz passband of 
the dual-band continuum setup, and comparing it with the maps 
obtained from the dual-band data.  Based on this cross-registration,
we show SiO maser maps, smoothed to 5.43 \kms\ resolution, superposed 
on the continuum emission of \OrionI\ in Fig.~\ref{fig:channel_maps}.
This figure shows that the continuum emission is precisely centered 
among the four SiO maser arms, whose innermost ends nestle tightly against 
the disk-like continuum structure.
 
\section {Results \& Discussion}

   Our images of the continuum emission from \OrionI\ at 43 GHz are 
shown in Fig.~\ref{fig:maps}.  The top and middle panels of the 
figure are maps made with resolutions of 50 and 34 mas, respectively.  
The total flux density of \OrionI\ is 13 mJy and the peak brightness 
at 34 mas resolution is 3.0~\mjyb.  
The emission appears to be composed of a compact component,
near the center of the source, and a component elongated NW--SE.  
Assuming the elongated component is approximately
uniformly bright, it would contribute about 0.8~\mjyb\ at the
center of the source, leaving 2.2~mJy for the compact component.
Subtracting a 0.8~mJy point-source centered at the 
position of peak brightness, we obtain the image shown in the 
bottom panel of Fig.~\ref{fig:maps}.  
This reveals a disk-like feature with
a radius of $\approx35$~AU and a brightness of about 1~\mjyb.  
Away from the center of the source, the true brightness is a lower limit, 
since the feature is not well resolved perpendicular to its elongation. 
We also note that the peak brightness along the disk-like feature 
does not follow a straight line on the sky.  Instead, it appears to bend as
might be expected from a warped disk.  

   In the following discussion, we assume that the compact emission 
comes from the immediate environment of a YSO, and that the elongated 
component traces a nearly edge-on disk, whose
spin axis is aligned northeast--southwest.   
Briefly, the evidence supporting this model is as follows.
{\citet{G03} report detection of a curved arc of SiO maser emission
bridging the gap between the base of the south and west arms.  
Evidence of this emission can also be seen in the 
$-3.47$ and $12.8$~\kms\ channel maps (Fig.~\ref{fig:channel_maps}).  
The bridge emission displays a radial velocity gradient, and some features have 
tangential proper motions, consistent with material rotating close to the 
nearside of a disk.  Such emission is not anticipated for a bipolar outflow.
Additionally, \citet{G03} note that \hho\ maser emission
comes from ``caps'' displaced predominantly 
$0\arcsper2$-$0\arcsper7$ toward the northeast and southwest of 
\OrionI, \ie along the disk spin axis.  These caps show outward
motion and indications of rotation; the red- and blue-shifted emission 
tend to lie on opposite sides of the spin axis, consistent with the inferred 
direction of disk rotation.
A more complete presentation of these findings will appear in 
\citet{G07}.

   At the center of the disk, the source appears slightly extended 
perpendicular to its elongation (\ie along ${\rm PA}=+45^\circ$).
It is possible that a weak jet emanates from the YSO, perpendicular 
to the disk, resulting in the extended appearance at the center.  
Clearly, higher sensitivity observations are needed to understand
this structure.

    What are the physical conditions in the \OrionI\ source?
To answer this question, one must know the emission mechanism 
(opacity source) for the cm-wave photons.  
The cm-to-mm wavelength spectrum of the entire source can be
characterized as a power law with flux density, $S_\nu$, rising 
with observing frequency, $\nu$, as $S_\nu \propto \nu^{1.6}$ 
\citep{MR95,B06}, approaching that of a black body.
Since the source is not well resolved spatially at
lower frequencies with the VLA, the spectral index does not 
allow us to discriminate between an inhomogeneous, single-component 
model (where the spectral index is shallower 
than 2.0, because unity optical-depth occurs at a smaller 
radius at higher frequencies) and a two-component model 
(with an optically thick central component and a partially
optically thin disk-like structure).

We think it unlikely that dust emission could be a dominant
contributor to the cm- to mm-wavelength emission of \OrionI.
A dense, warm, dusty disk would be expected
to show a plethora of molecular lines at mm/sub-mm wavelengths.
While \citet{B06} find numerous, strong, molecular lines toward the 
nearby ``hot core,'' they find no strong lines toward the position of 
\OrionI\ (only weak SO lines and, of course, the strong SiO masers
slightly offset from \OrionI).  Thus, we look to other emission 
mechanisms to explain both the YSO peak and the elongated disk 
components.

The observations could be modeled with gas at $\approx8000$~K, 
where hydrogen is fully ionized (proton-electron bremsstrahlung).
In this case, the data require an optically thick central component and 
a partially optically thin disk component.  Alternatively, the emission 
could be partially optically-thick from gas at $<5000$~K, where 
hydrogen is predominantly neutral and free electrons come from low 
ionization-potential metals (H-minus free-free).  
The latter case applies in the ``radio photospheres'' of Mira 
variables at roughly 2 stellar radii \citep{RM97}. 
In the following subsections, we present two 
classes of models for the \OrionI\ cm-wave emission.
These models are exploratory and designed only to elucidate 
characteristic physical conditions.

\subsection {Collisional Ionization: \Hminus\ Opacity}

    In several ways \OrionI\ appears similar to a Mira-like variable 
star.  Such stars display OH, \hho, and SiO masers, as seen in \OrionI.  
In addition, Mira variables have continuum emission detectable with 
the VLA at cm wavelengths, with brightness temperatures of 
$\approx1600$~K \citep{RM97}.  The radio continuum of
Mira variables occurs in an optically-thick (spectral index $\approx1.9$)
``radio photosphere'' with characteristic temperature of $\approx1600$~K 
and density of $\sim10^{12}$ \percmcube.
Under these conditions the dominant opacity source is H-minus free-free
interactions, coming from free electrons interacting with neutral 
hydrogen (either atomic or molecular).   
This is analogous to normal proton-electron bremsstrahlung, 
except that the interaction is about $10^4$ weaker \citep{DL66},
requiring correspondingly higher densities.
At these temperatures and densities, sufficient free electrons can 
be created by collisional ionization of Na and K \citep{RM97}.  

The \sioone\ maser emission at 43 GHz originates from the first
vibrationally-excited state at $\approx1800$~K above the ground-state,
and models of maser pumping require temperatures of $\approx1200$~K and
hydrogen densities of $\sim10^{9-10}$ \percmcube\ for strong maser 
action \citep{ABG89,LE92,B94}.
Since the continuum emission region is more compact and has a 
higher (brightness) temperature than required for SiO maser 
excitation, finding the loci of SiO maser emission extending outward 
from the continuum, as shown in Fig.~\ref{fig:channel_maps}, is 
reasonable and as observed for Mira-like variables \citep{RM03}.  
As both Miras and \OrionI\ display similar cm-to-mm wavelength 
spectral indexes and have a similar configuration of continuum and 
maser emission, there is circumstantial evidence for similar 
physical conditions and mechanisms.

    We have explored models for the disk-like emission of Source-I 
owing to \Hminus\ opacity. Assuming conditions similar to a Mira-like 
radio photosphere, material at density  
$\sim10^{11}$~\percmcube\ and temperature $\approx1500$~K  
requires a path length of $\approx2$ AU in order to achieve an 
optical depth of $\approx0.5$ \citep{RM97}, possibly explaining the
observed disk brightness temperature of $\gax450$~K.  
This path length is about 10\% of the disk radius and, thus, is
easily achievable.  Such a model has the benefit
that a single power-law can explain the observed spectral energy
distribution for the entire \OrionI\ source between 8 and 350 GHz.
However, recently \citet{B06} have measured the flux density of 
\OrionI\ at 690~GHz to be between 3.5 and 9.9~Jy.  
Since the extrapolation of the cm-wave continuum spectrum to 
690~GHz predicts under 2~Jy, an additional component (\eg dust on
a scale of 0.2\arcs\ to 2\arcs) 
seems required to explain the sub-mm wavelength spectrum of \OrionI, 
making a single emission mechanism unlikely.

   For \OrionI, we observe a disk-like component that extends 
to about $0.08\arcs$ ($\approx40$~AU) from the star.   
We have attempted to model the brightness profile of such a disk 
in a manner similar to that done for the radio photospheres of Miras 
\citep{RM97}, but with a disk geometry.  
Specifically, we assume an edge-on disk that is
centrally heated and is optically thick to most of the
radiation from the YSO.  In Fig.~\ref{fig:slice_fits}
we plot the observed brightness temperature in the map
with 34 mas resolution (middle panel of Fig.~\ref{fig:maps}) 
as a function of position along the disk elongation.  
The physical parameters of the central star and disk are 
listed in Table~\ref{table:fit_parameters} for model A.
In Fig.~\ref{fig:slice_fits} we overplot the model brightness 
(blue dotted line) convolved with the observed restoring beam.  
While model A provides a reasonable fit to the observations, the
model requires the central star to have a total luminosity 
of $\sim3\times10^5$~\Lsun\ and a disk mass of $\sim3$~\Msun.
These are general characteristics of this class of models and
are not sensitive to details of the parameters.
Note that a similarly large luminosity may also be required
to explain the SiO masers \citep{MR95}.
Reducing the stellar luminosity requires substantially increasing
the disk mass (in order to increase opacity and maintain a high disk 
brightness temperature).
Thus, such a model requires a fairly massive disk and a luminosity 
exceeding that from the IRc~2 region \citep{Gez98,G04} and, perhaps, 
even that of the entire \OrionKL\ nebula \citep{T86}.  
Since other energetic sources exist nearby (\eg  Source-n), 
we conclude that the disk component of \OrionI\ probably is not 
thermally (collisionally) excited by a central source.

While central heating of the disk component (and also the SiO masers)
may be ruled out on energetic grounds, the material in the disk
and the SiO masers may be partially locally heated by accretion
processes.  Dissipation of energy in the disk  
could raise the temperature above that allowed by radiative
equilibrium with the central star.  Since the volume of the disk
can be considerably less than that of a sphere of the same radius,
increasing the disk temperature in this manner can require less
total energy than for central heating alone.  Thus, we evaluated
models that allowed the disk temperature to fall with radius, $r$, 
more slowly than $r^{-1/2}$.   One such model, described in 
Table~\ref{table:fit_parameters} as model B and shown in 
Fig.~\ref{fig:slice_fits}, fits the data well and requires a
central star luminosity of $5 \times 10^4$~\Lsun, comfortably
below the observational limits.
  
As pointed out by \citet{MR95}, the SiO maser excitation also
requires a very high luminosity source, were it to be centrally
heated (\ie assuming radiative equilibrium: $L = \sigma T^4 4\pi r^2$).  
However, infalling material might interact with
outflowing material (and possibly magnetic fields) in the 
conical walls of a bi-polar outflow.  This may add heat, augmenting the
central source and providing the necessary high temperatures
($\approx1200$~K) for strong SiO maser emission.

\subsection {Photo-ionization: \pe\ Bremsstrahlung}

  Given the high luminosities and disk masses characteristic
of models involving thermal ionization and \Hminus\ opacity, we now
consider a hotter central star.  The observed brightness can be
modeled with an early B-type star, which photo-ionizes a 
moderate density plasma.   Indeed, since proton-electron bremsstrahlung
is $\sim10^4$ times stronger than \Hminus\  per interaction,
the disk plasma need only have a density $\sim10^{-4}$ times
lower.

   Assume that \OrionI\ contains a hot central star and a 
photo-ionized disk (or a photo-ionized surface layer).   
The disk may contain a neutral central layer, which provides a 
reservoir for material that can be photo-ionized by the YSO \citep{H94}.  
The sub-mm wave spectrum of \OrionI,  measured
by \citet{B06}, suggests a dust component dominates above 300 GHz,
leaving a bremsstrahlung spectrum that becomes  
optically thin above $\approx100$~GHz.  Such a turnover frequency 
can come from an electron density of $\sim10^7$\percmcube\ 
over a path length of 35 AU, comparable to the observed radius.  
These parameters yield an excitation parameter, $U$, 
of $\sim10$~pc\percmsquare, which could come from a ZAMS B0--B1 star 
\citep{P73} of $\approx10$~\Msun\ and a luminosity 
approaching $10^4$~\Lsun. A star of $\gax6$~\Msun\ is consistent with 
the rotation and expansion seen in VLBA maps of the SiO masers
\citep{G03,C05}.

   In Fig.~\ref{fig:slice_fits} (dashed green line), 
we present a simple model of a brightness temperature 
profile along an edge-on, photo-ionized disk with a constant temperature.  
The physical parameters of the star and disk are given in the 
Table~\ref{table:fit_parameters} as model C.
This model provides a reasonable fit to the data, demonstrating
that a photo-ionized disk can explain our observations.

   Recently, \citet{K02,K03} and \citet{KW06} have shown that the 
inner portion of a disk can be fully ionized and still allow for 
continued accretion onto a massive protostar.  They point out
that inside a critical radius $r_c = GM/2c_s^2$, where 
$G$ is the gravitational constant, $M$ is the mass of the
central protostar, and $c_s$ is the sound speed in the
(neutral or ionized) material, the protostar's gravity exceeds 
the thermal pressure.  For the stellar parameters given
above for a ZAMS B0--B1 star, ionized accretion
can proceed inside of the critical radius of about 25~AU.
This critical radius is similar to the 35~AU radius of the
disk observed at 43~GHz in \OrionI.

   \citet{K07} explored models of ionized accretion in the
presence of significant angular momentum.  The example 
shown in the {\it right-hand} panel of his Fig.~1 corresponds
to a star of 20~\Msun, an ionizing flux of $3\times10^{46}$
photons~s$^{-1}$ (approximately a B0 -- 09.5 star), and
accreting material with specific angular momentum of 
0.16~\kms~pc.  The resulting critical radius for accretion of ionized 
gas is 54~AU.  Scaling this to a 10~\Msun\ star gives a critical
radius of 27~AU, reasonably consistent with our observations.

An ionized accretion disk offers a natural explanation for the dearth 
of sub-mm wavelength spectral-lines, observed by \citet{B06} toward \OrionI,
that would otherwise be expected for a dense and warm neutral disk.   
Thus, \OrionI\ may be a good example of a massive YSO accreting 
material after ionizing its inner accretion disk.  
If the disk is maintained at a temperature greater than $\approx1500$~K
out to a radius of 35 AU, as we observe in \OrionI, this may preclude 
planetary formation.  Planet formation is thought to occur on a much
longer time scale than the formation of a massive star.
In order to form planets around massive stars, 
the early phases of planetesimal formation would have to overcome 
temperatures high enough to sublimate dust, and certainly not conducive 
to rapid grain growth.  This difficulty is in addition to the well known
problem of the short timescales for the formation of massive stars and
dispersal of their disks. 

\section{Other High-mass Protostars}

While \OrionI\ is probably the nearest high-mass protostellar 
object, more distant candidates include CRL 2136, W 33, AFGL 2591, 
and NGC 7538/IRS 9.  These objects display cm-to-mm wavelength spectra
closely resembling that of \OrionI, and whose continuum emissions
are unresolved (or only marginally resolved) at 40 mas resolution 
\citep{M04,vdT05}.
While none of these candidates have been observed to have
SiO maser emission, \citet{M04} find that CRL 2136 has \hho\ 
masers very close (in projection) to the continuum emission;  
the masers might arise in dense, hot gas following an accretion shock.

\section{Future Observations}

   Our current image of \OrionI, at a resolution of 34~mas 
($\approx 16$ AU), appears to show an ionized disk 
around a massive YSO.  
While this may be the best image to date of such a disk,
our data are limited both in sensitivity and angular resolution.  
We have only about five resolution elements along
the disk, and we have not clearly resolved the emission perpendicular
to the disk.   In order to improve significantly the sensitivity,
of this image, we probably must await the completion of the EVLA
phase-I project.  Improved angular resolution, 
with the required higher sensitivity, could be achieved with the 
planned increase in baseline length of the EVLA phase-II 
or, in the long-term, with the Square Kilometer Array.

\acknowledgments  
We thank L. Matthews for suggestions to improve the manuscript.

\clearpage

\begin{deluxetable}{lccccc}
\tabletypesize{\scriptsize}
\tablecaption   {Star \& Disk Model Parameters \label{table:fit_parameters} }

\tablehead{ \colhead{Parameter} &\colhead{Units} &\colhead{~~~~~~~~~~~~~~} 
            &\colhead{Model A} &\colhead{Model B} &\colhead{Model C}
          }
\tablecolumns{6}
\startdata
Opacity source ....          &           &&\Hminus &\Hminus &\pe\ bremsstrahlung\\
\\
Stellar Radius ($R_*$) ....  &AU         &&4      &4        &0.25         \\
Disk Temperature at  $R_*$ ..  &K        &&4500   &3000     &8000         \\
Disk Density at $R_*$  ....  &\percmcube &&$1\times10^{14}$ &$1\times10^{14}$ &$6\times10^{10}$ \\ 
Disk Thickness         ....  &AU         &&25     &25       &6          \\
Temperature power law index ..&          &&$-0.5$ &$-0.33$  &$+0.0$       \\
Density power law index ...   &          &&$-2.0$ &$-2.00$  &$-2.0$       \\
\\
Stellar Luminosity ...        &\Lsun     &&$3\times10^5$ &$5\times10^4$ &$1\times10^4$ \\
Disk Mass...                  &\Msun     &&$3$    &$3$      &... \\
\enddata
\tablecomments{\pe\ Bremsstrahlung model assumes a constant electron 
temperature and is insensitive to the stellar radius; also while the
ionized disk mass is $<<1$~\Msun, it is possible to have a substantial
neutral mass that does not contribute to the emission.  
Power law indexes are defined as $\propto r^{{\rm index}}$.
              }
\end{deluxetable}

\clearpage

\begin{figure}
\epsscale{0.40}
\plotone{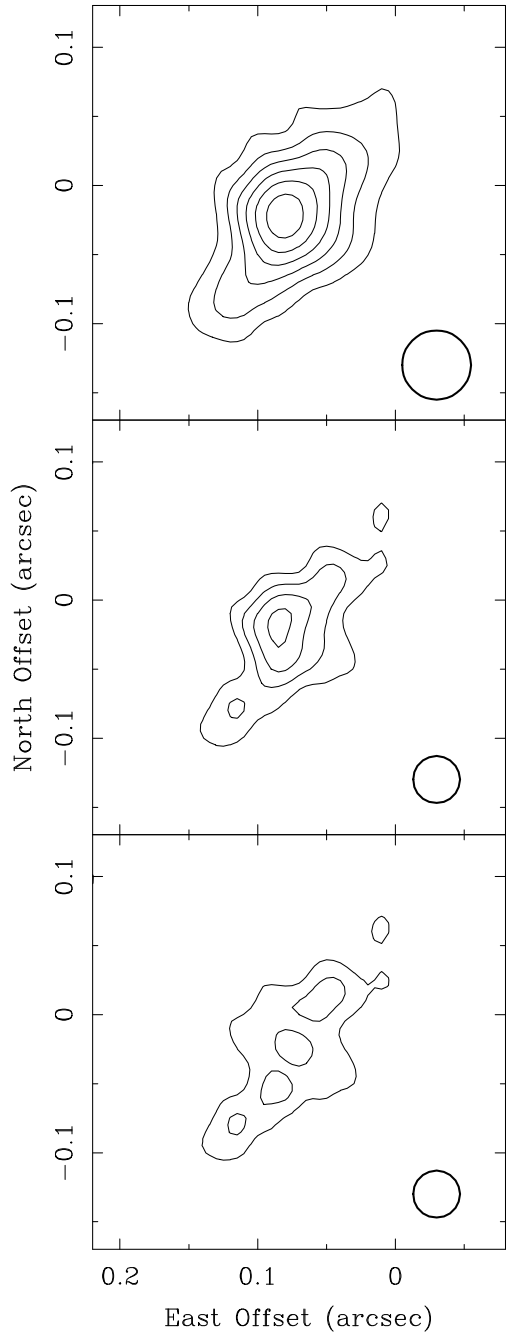}
\caption{Continuum images of \OrionI\ at 43 GHz
made with the VLA in the A-configuration.  
The image in the {\it upper panel} is at 50~mas (FWHM) resolution 
and the image in the {\it middle panel} is at 34~mas resolution.  
The emission is elongated NW--SE and may
be from an disk surrounding a massive YSO.
The brightest component near the center of the disk may be
unresolved.  When we subtract a point-like 2.2~mJy component
from the (13~mJy total) emission, we obtain the image in
the {\it lower panel.}   In all images the contour levels are
at integer multiples of 0.5~\mjyb.  The FWHM of the restoring 
beams are shown in the bottom right corner of each panel.
At a distance of 480~pc, 0.1\arcs\ corresponds to 48~AU.
   \label{fig:maps}
        }
\end{figure}

\clearpage

\begin{figure}
\epsscale{0.9}
\plotone{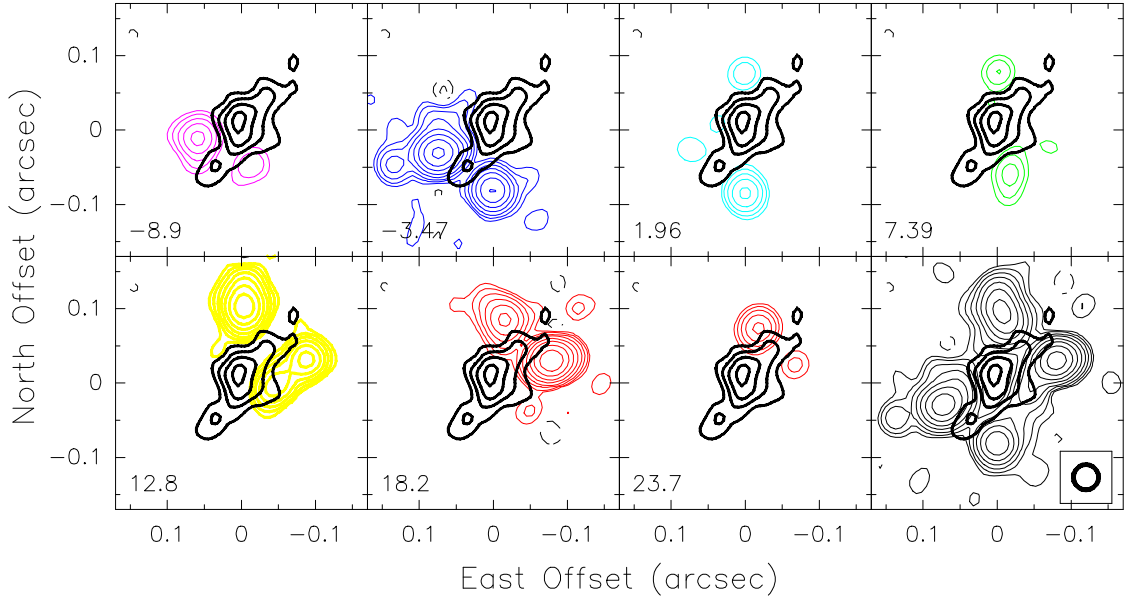}
\caption{\OrionI\ \sioone\ maser channel maps ({\it colored contours}) 
superposed on the 34~mas resolution continuum emission 
({\it heavy black contours}).  The colors are chosen to approximate those 
used in Fig. 1a of \citet{G98}; contouring levels start at 3~\jyb\ 
and increase by factors of 2. Center LSR velocities of the
5.43~\kms\ wide channels are given in the lower left corner of
each panel.  Continuum contours are integer multiples of 0.5~\mjyb.
The {\it bottom right panel} shows a map of the integrated SiO emission,
with light contours at integer multiples of 0.18 Jy~beam$^{-1}$\kms. 
All position offsets are relative to \OrionI, whose position is
$(\alpha,\delta)_{\rm J2000}=(05^{\rm h}35^{\rm m}14.5121^{\rm s},
~-05^\circ22\arcm0.521\arcs)$
at 2000 Nov.~13 \citep{G05}.
   \label{fig:channel_maps}
        }
\end{figure}

\clearpage

\begin{figure}
\epsscale{0.65}
\plotone{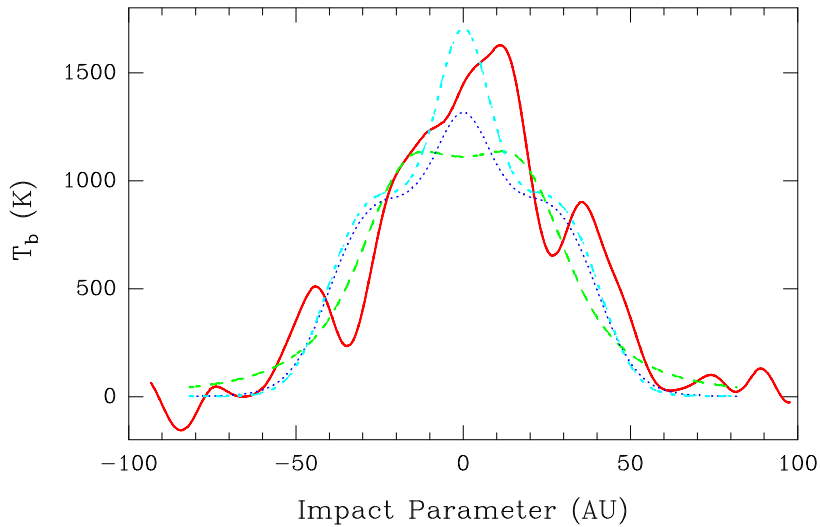}
\caption{Brightness temperature profiles of the 43~GHz continuum 
emission of \OrionI.
The observed profile ({\it solid red line}) is along the elongation of 
the edge-on disk-like emission in the 34~mas resolution image.
Model profiles are shown for an edge-on disk with collisional ionization
for central heating only (Model A, {\it blue dotted line})
and with additional local heating 
(Model B, {\it cyan dash-dotted line}; see \S3.1).
A photo-ionized disk model is also shown
(Model C, {\it green dashed line}; see \S3.2).  Parameters for all models 
are listed in Table~\ref{table:fit_parameters}, and the model brightnesses
have been convolved with a 34~mas Gaussian to approximate the observations.
   \label{fig:slice_fits}
        }
\end{figure}


\begin{thebibliography}{}
\bibitem[Alcolea, Bujarrabal \& Gallego (1989)]{ABG89}
   Alcolea, J., Bujarrabal, V. \& Gallego, J. D. 1989,
   \aaps, 211, 187
\bibitem[Baars \etal (1977)]{B77}
   Baars, J. W. M., Genzel, R., Pauliny-Toth, I. I. K \&
   Witzel, A. 1977, \aaps, 61, 99 
\bibitem[Beuther \etal(2006)]{B06}
   Beuther, H. \etal 2006, \apj, 636, 323
\bibitem[Bujarrabal (1994)]{B94}
   Bujarrabal, V. 1994, \aaps, 285, 953
\bibitem[Cohen \etal (2006)]{C06}
   Cohen, R. J., Gasiprong, N., Meaburn, J. \& Graham, M. F.
   2006, \mnras, 367, 541
\bibitem[Cunninghan, Frank \& Hartmann (2005)]{C05}
   Cunningham, A., Frank, A. \& Hartmann, L. 2005, \apj, 631, 1010
\bibitem[Dalgarno \& Lane (1966)]{DL66}
   Dalgarno, A. \& Lane, N. F. 1966, \apj, 145, 623
\bibitem[Dougados \etal (1993)]{D93}
   Dougados, C, L\'ena, P., Ridgway, S. T., Christou, J. C. \&
   Probst, R. G. 1993, \apj, 406, 112
\bibitem[Doeleman, Lonsdale \& Pelkey (1999)]{D99}
   Doeleman, S. S., Lonsdale, C. J. \& Pelkey, S. 
   1999, \apj, 510, L55
\bibitem[Downes \etal (1981)]{D81}
   Downes, D., Genzel, R., Becklin, E. E. \& Wynn-Williams, G. C. 
   1981, \apj, 244, 869
\bibitem[Genzel \etal (1981)]{G81}
   Genzel, R., Reid, M. J., Moran, J. M. \& Downes, D. 1981,
   \apj, 244, 884
\bibitem[Gezari (1992)]{Gez92}
   Gezari, D. Y. 1992, \apj, 396, L43
\bibitem[Gezari, Backman \& Werner (1998) ]{Gez98}
   Gezari, D. Y., Backman, D. E. \& Werner, M. W. 1998, \apj, 509, 283
\bibitem[G\'omez \etal (2005)]{G05}
   G\'omez, L., Rodr\'iguez, L. F., Loinard, L., Lizano, S., 
   Poveda, A. \& Allen, C. 2005, \apj, 635, 1166
\bibitem[Greenhill \etal (1998)]{G98}
   Greenhill, L. J., Gwinn, C. R., Schwartz, C., Moran, J. M.
   \& Diamond, P. J. 1998, \nat, 416, 59 
\bibitem[Greenhill \etal (2004a)]{G03}
   Greenhill, L. J., Chandler, C. J., Reid, M. J., Diamond, P. J.
   \& Moran, J. M. 2004a, in ``Star Formation at High Angular 
   Resolution,'' Proceedings of IAU Symposium 221, eds. M. G. Burton,
   R. Jayawardhana \& T. L. Bourke, (San Francisco: ASP), p. 155 
\bibitem[Greenhill \etal (2004b)]{G04}
   Greenhill, L. J., Gezari, D. Y., Danchi, W. C., Najita, J.,
   Monnier, J. D. \& Tuthill, P. G. 2004b, \apj, 605, L57
\bibitem[Greenhill \etal (2007)]{G07}
   Greenhill, L. J., Chandler, C. J., Reid, M. J., Diamond, P. J.
   \& Moran, J. M. 2007, in preparation
\bibitem[Hasegawa \etal (1985)]{H85}
   Hasegawa, T. \etal 1985, in ``Masers, Molecules and Mass Outflows
   in Star Forming Regions,'' ed. A. Haschick 
   (Haystack Obs., Westford, MA), p. 275
\bibitem[Hollenbach \etal (1994)]{H94}
   Hollenbach, D., Johnstone, D., Lizano, S. \& Shu, F. 1994,
   \apj, 428, 654
\bibitem[Keto (2002)]{K02}
   Keto, E. 2002, \apj, 580, 980
\bibitem[Keto (2003)]{K03}
   Keto, E. 2003, \apj, 599, 1196
\bibitem[Keto (2007)]{K07}
   Keto, E. 2007, submitted to \apj
\bibitem[Keto \& Wood (2006)]{KW06}
   Keto, E. \& Wood, K. 2006, \apj, 637, 850
\bibitem[Kleinmann \& Low (1967)]{KL67}
   Kleinmann, D. E. \& Low, F. J. 1967, \apj, 149, L1
\bibitem[Lockett \& Elitzur (1992)]{LE92}
   Lockett, P. \& Elitzur, M. 1992, \apj, 399, 704
\bibitem[Menten \& Reid(1995)]{MR95} 
   Menten, K. M.~\& Reid, M. J.\ 1995, \apj, 445, L157
\bibitem[Menten \& van der Tak(2004)]{M04} 
   Menten, K. M. \& van der Tak, F. F. S. 2004, \aap, 414, 289
\bibitem[Morino \etal (1998)]{M98}
   Morino, J.-I., Yamashita, T., Hasegawa, T. \& Nakano, T.\ 
   1998, \nat, 393, 340
\bibitem[Panagia (1973)]{P73}
   Panagia, N. 1973, \aj, 78, 929
\bibitem[Plambeck \etal (1995)]{P95}
   Plambeck, R. L., Wright, M. C. H., Mundy, L. G. \& Looney, L. W.
   1995, \apj, 455, L189
\bibitem[Reid \& Menten (1997)]{RM97}
   Reid, M. J. \& Menten, K. M. 1997, \apj, 476, 327 
\bibitem[Reid \& Menten (2003)]{RM03}
   Reid, M. J. \& Menten, K. M. 2003, in 
   Mass-losing pulsating stars and their circumstellar matter, 
   Astrophysics and Space Science Library, Vol. 283, 
   eds. Y. Nakada, M. Honma and M. Seki. 
   (Kluwer Academic Publishers, Dordrecht), p. 283
\bibitem[Shuping, Morris \& Bally (2004)]{S04}
   Shuping, R. Y., Morris, M. \& Bally, J. 2004, 
   \apj, 128, 363
\bibitem[Thronson \etal (1986)]{T86}
   Thronson, H. A. \etal 1986, \aj, 91, 1350
\bibitem[van der Tak \& Menten(2005)]{vdT05} 
   van der Tak, F. F. S. \& Menten, K. M. 2005, \aap, 437, 947

\end{thebibliography}
\end{document}